\documentclass[amsmath,amssymb,aps,pra,twocolumn,superscriptaddress,floatfix]{revtex4-1}

\usepackage{graphicx}
\usepackage[version=3]{mhchem}
\usepackage{multirow}
\usepackage{epstopdf}
\usepackage{epsfig}

\graphicspath{{Figures/}}

\begin{document}
\setlength{\parskip}{0pt}
\setlength{\parsep}{0pt}

\title{Measurement of low--energy \ce{Na+}-- \ce{Na} total collision rate in an ion--neutral hybrid trap}
\author{D.~S.~Goodman}
\affiliation{Department of Physics, University of Connecticut, Storrs, Connecticut 06269}
\author{J.~E.~Wells}
\affiliation{Department of Physics, University of Connecticut, Storrs, Connecticut 06269}
\author{J.~M.~Kwolek}
\affiliation{Department of Physics, University of Connecticut, Storrs, Connecticut 06269}
\author{R.~Bl\"{u}mel}
\affiliation{Department of Physics, Wesleyan University, Middletown, Connecticut 06459}
\author{F.~A.~Narducci}
\affiliation{Naval Air Systems Command, EO Sensors Division, Bldg 2187, Suite 3190 Patuxent River, Maryland 20670, USA }
\author{W.~W.~Smith}
\affiliation{Department of Physics, University of Connecticut, Storrs, Connecticut 06269}
\date{\today}

%%%%%%%%%%%%%%%%%%%%%%%%%%%%%%%%%%%%
\begin{abstract}
We present measurements of the total elastic and resonant charge-exchange ion-atom collision rate coefficient $k_\mathrm{ia}$ of cold sodium (\ce{Na}) with optically-dark low energy \ce{Na+} ions in a hybrid ion-neutral trap. To determine $k_\mathrm{ia}$, we measured the trap loading and loss from both a \ce{Na} magneto-optical trap (MOT) and a linear radio frequency quadrupole Paul trap.  We found the total rate coefficient to be $7.4 \pm 1.9 \times 10^{-8}$ cm$^3$/s for the type I \ce{Na} MOT immersed within an $\approx 140$ K ion cloud and $1.10 \pm 0.25 \times 10^{-7}$  cm$^3$/s for the type II \ce{Na} MOT within an $\approx 1070$ K ion cloud.  Our measurements show excellent agreement with previously reported theoretical fully quantal \textit{ab initio} calculations.  In the process of determining the total rate coefficient, we demonstrate that a MOT can be used to probe an optically dark ion cloud's spatial distribution within a hybrid trap.
\end{abstract}
\pacs{}
\maketitle
%%%%%%%%%%%%%%%%%%%%%%%%%%%%%%%%%%%%
%%%%%%%%%%%%%%%%%%%%%%%%%%%%%%%%%%%%
\section{INTRODUCTION}
\label{sec:Introduction}
\vspace{-0.5em}
A hybrid ion-neutral trap is a combination of two normally separate technologies -- a cold neutral atom trap within an ion trap, e.g., a linear radiofrequency quadrupole or octupole \cite{Deiglmayr:2012} Paul trap (LPT). Typically, the neutral trap consists of a magneto-optical trap (MOT) \cite{Sivarajah:2012, RaviAPB:2012,Sullivan:2011,Hall:2011}, an optical dipole trap (ODT) \cite{Haze:2013}, or a magnetically trapped Bose-Einstein condensate (BEC) \cite{ZipkesNature:2010,Schmid:2010}.  Recently, a hybrid trap was developed that also incorporates an optical cavity \cite{Ray:2014}. Over the past decade, since the hybrid trap was originally proposed \cite{Smith:2003,Smith:2005}, both experimental \cite{Rellergert:2011, Sullivan:2011, Sullivan:2012, Grier:2009, ZipkesPRL:2010, ZipkesNature:2010, Schmid:2010, Hall:2011, Ravi:2012, Lee:2013, Ray:2014, Chen:2014, Cetina:2012, Goodman:2012, Sivarajah:2012, Smith:2014, Haze:2013, Ratschbacher:2012} and theoretical \cite{Cote:2000, Makarov:2003, Zhang:2009, Hudson:2009, McLaughlin:2014, Idziaszek:2009, Li:2012, Tacconi:2011} interest in low-energy ion-neutral collisions has surged.

Cold ion-neutral collisions are of intermediate range between neutral-neutral and ion-ion. Compared to neutral-neutral van der Waals cross sections $\sim 1$ a.u., they have large elastic scattering cross sections $\sim10^6$ a.u.~at 1 mK \cite{Cote:2000,Makarov:2003,Zhang:2009}. These large cross sections are a consequence of the ion polarizing the colliding neutral atom, which leads to universal long-range polarization potentials \cite{Krems:2009}, with the principal term $\propto -C_4 /R^4$. Here $C_4$ is the atomic dipole polarizability of the neutral collision partner and $R$ is the internuclear ion-atom separation.

The large ion-neutral elastic scattering cross sections have been utilized to demonstrate hybrid trap sympathetic cooling \cite{Smith:2003,Smith:2005,Goodman:2012,Hudson:2009} of atomic ions' translational motion \cite{Sivarajah:2012,Ravi:2012,Deiglmayr:2012,Ray:2014} and molecular ions' internal degrees-of-freedom \cite{Rellegert:2013}.  Additionally, there have been several measurements of low-energy ion-neutral elastic \cite{ZipkesNature:2010,Schmid:2010,Haze:2013} and charge-exchange \cite{Rellergert:2011, Sullivan:2012, Grier:2009,  Hall:2011, Hall:2013} rate coefficients within hybrid traps. The rate coefficient measurments are of interest to both astrophysics  \cite{Dalgarno:2004,Balakrishnan:2001,Stancil:1996,Kharchenko:2001,Kirby:1995} and quantum information \cite{Ratschbacher:2012, DeMille:2002}.

Several methods have been used to measure scattering rates using a hybrid trap, including monitoring the neutral atom fluorescence decay from an ODT \cite{ZipkesNature:2010,Schmid:2010,Haze:2013} and measuring the ion fluorescence decay from a Paul trap \cite{Grier:2009, Rellergert:2011}. Recently, hybrid trap measurements of the total elastic and change-exchange collision rate for closed shell, optically dark \ce{Rb+} ions on \ce{Rb} (rubidium) were reported by Lee \textit{et al}.~\cite{Lee:2013}. The fluorescence of the neutral species is used to measure the total collision rate of optically dark ions. Our method uses the loading and decay of both the atoms in the MOT and the dark ions in the LPT to determine the collision rate. Additionally, for optically accessible ions, the methods presented here for determining the total collision rate can be used in conjunction with the previously demonstrated charge-exchange measurement methods to isolate the elastic collision rate.

We present measurements of the total collision rate coefficient for the \ce{Na+}-- \ce{Na} (sodium) system. Our experimental results show excellent agreement with previously reported fully quantal \textit{ab intio} theoretical calculations \cite{Cote:2000}.  We use a similar experimental procedure to the one reported in Ref.~\cite{Lee:2013}. However, we find deviations between our experimental results and the LPT loading model presented in Ref.~\cite{Lee:2013}.

This manuscript is organized as follows:  In Sec.~\ref{sec:Background} we begin with a discussion of our hybrid apparatus, the semiclassically predicted \ce{Na+}-- \ce{Na} total collision rate model, and our experimental model.  In Sec.~\ref{sec:ExpResults} we present our results for MOT loading, LPT loading, and determining the volume of the optically dark \ce{Na+} ion cloud. we conclude in Sec.~\ref{sec:Con}
\vspace{-1em}
%%%%%%%%%%%%%%%%%%%%%%%%%%%%%%%%%%%%
%%%%%%%%%%%%%%%%%%%%%%%%%%%%%%%%%%%%
\section{Background}
\label{sec:Background}
\vspace{-1em}
%%%%%%%%%%%%%%%%%%%%%%%%%%%%%%%%%%%%
\subsection{Apparatus}
\label{sec:Apparatus}
\vspace{-1em}
\subsubsection{Magneto-optical trap}
\label{sec:MOT}
\begin{figure}[t]
   \centering
   \includegraphics[width=3.25in]{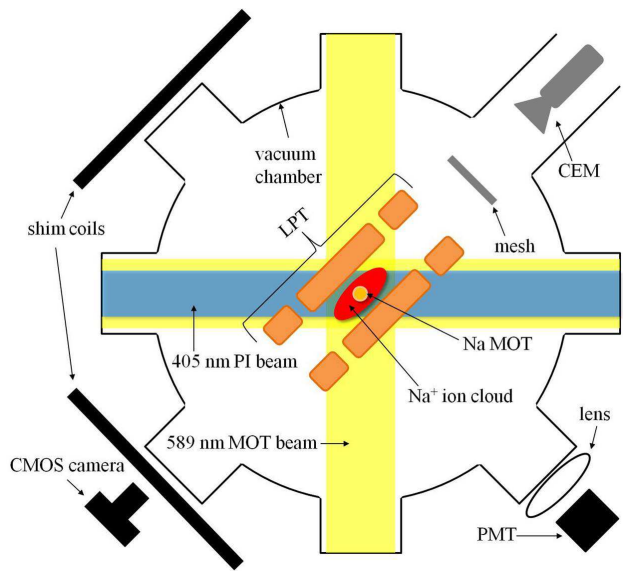}
   \caption{(Color online). Schematic of the hybrid trap apparatus within the vacuum chamber (top view, not to scale).  The \ce{Na} magneto-optical trap (MOT) is created in the center of the segmented linear Paul trap (LPT) with six 589 nm laser beams. MOT fluorescence measurements can be taken with our CMOS camera or with our photomultiplier tube (PMT). The MOT can be translated within the LPT using the electromagnet shim coils. A pair of anti-Helmholtz coils and a third shim coil sit outside the vacuum chamber and are not shown in the figure. The LPT is loaded with \ce{Na+} ions by photoionization (PI) of excited \ce{Na}($3\mathrm{P}$) MOT atoms with a 405 nm laser beam collinear with one MOT beam.  Ions are destructively detected by extracting the ions toward the Channeltron electron multiplier (CEM).  The mesh is both used for extraction ion optics and to create a more uniform gain across the CEM cone \cite{Channeltron:2}.}
   \label{fig:ap}
\vspace{-0.75em}
\end{figure}

\vspace{-1em}
A description of our experimental apparatus can also be found in our earlier references \cite{Goodman:2012, Sivarajah:2012}, but for the convenience of the reader we will briefly describe the apparatus here. A diagram of the apparatus can be found in Fig.~\ref{fig:ap}.

Our group's hybrid trap consists of a standard vapor-cell \ce{Na} MOT \cite{Raab:1987,Monroe:1990,Wipple:2001} concentric within a segmented LPT \cite{Prestage:1989,Drewsen:2000} and held in a vacuum chamber at a pressure $\sim 10^{-10}$ Torr.  The MOT is loaded with $346 \pm 3$ K \ce{Na} vapor from a biased getter source within the vacuum chamber.  The MOT simultaneously uses velocity and spatially dependent light pressure forces that both damp and trap the neutral \ce{Na} atoms \cite{Raab:1987}. This force is provided by three pairs of counterpropagating circularly polarized 589 nm laser beams intersecting within an added quadrupole magnetic field gradient of $\approx 30~\mathrm{Gauss}/\mathrm{cm}$, created with external anti-Helmholtz electromagnet coils. The 589 nm radiation is frequency stabilized to the saturation absorption spectrum of a \ce{Na} vapor cell. Additionally, three shim coils (two of them shown in Fig.~\ref{fig:ap}) are used for translating the MOT location within the LPT.  We have seen no experimental evidence to suggest that the MOT apparatus interferes with the operation of the LPT apparatus or vice versa \cite{Sivarajah:2012,RaviAPB:2012,Sullivan:2011}.

\begin{figure}[t]
   \centering
   \includegraphics[width=2.6in]{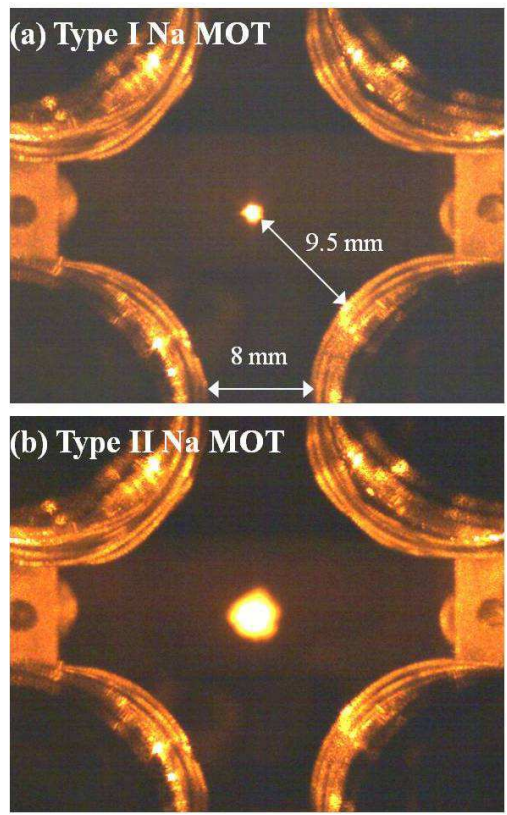}
   \caption{(Color online). CMOS camera image, without false coloring, of the smaller, denser, and colder type I MOT (a) and the larger, warmer type II MOT (b).  As depicted in Fig.~\ref{fig:ap}, the camera is looking down the axial (long) dimension of the segmented LPT. The inner edges of the LPT's end-segment electrodes can be seen in the corners of the image. The images are saturated to be more visually striking, but care is taken to avoid saturation when taking data.}
   \label{fig:MOTs}
\vspace{-0.75em}
\end{figure}

\begin{figure}[t]
   \centering
   \includegraphics[width=3.25 in]{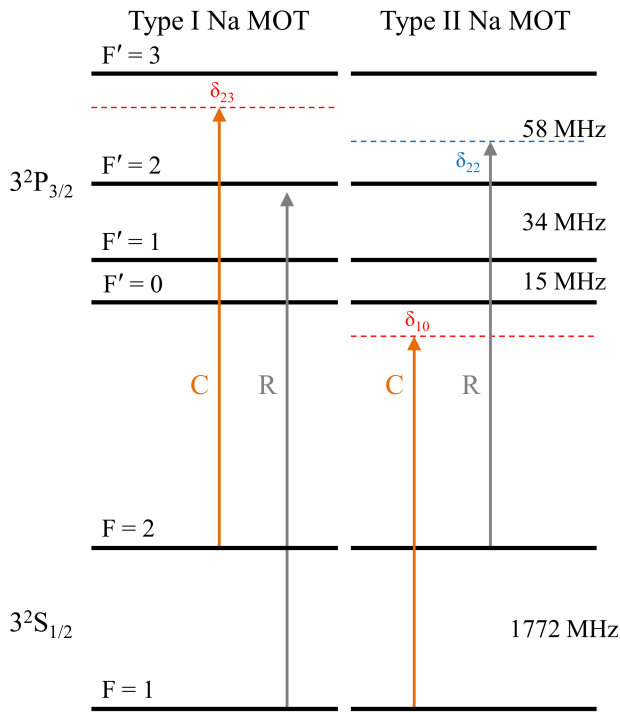}
   \caption{(Color online). Energy level diagram that shows the hyperfine structure of the D2 \ce{Na} line. We indicate the cooling laser (C) and the repumping laser (R) transitions for the type I and II MOTs. Each detuning from resonance ($\delta_{23}$ and $\delta_{22}$) was chosen for maximum MOT fluorescence intensity. For the type I MOT, the cooling laser is the carrier signal from our electro-optical modulator (EOM).  For the type II MOT, the cooling laser is the EOM sideband signal.}
   \label{fig:NaLvl}
\vspace{-0.75em}
\end{figure}

We can create two \ce{Na} MOTs that use different hyperfine cycling transitions: type (I) 3$\mathrm{S}$ $\mathrm{F}=2 \rightarrow 3\mathrm{P}\; \mathrm{F}^\prime=3$ or type (II) $3\mathrm{S}\; \mathrm{F}=1 \rightarrow 3\mathrm{P}\; \mathrm{F}^\prime= 0,1$ \cite{Tanaka:2007}.  Images taken with our CMOS camera of both MOTs (looking down the LPT's long axis) are shown in Fig.~\ref{fig:MOTs}.  A diagram of the energy level structure and laser cooling schemes for both MOTs is shown in Fig.~\ref{fig:NaLvl}.

By adjusting the relative MOT cooling beam intensities, both MOTs can be formed with approximately spherical Gaussian spatial distributions, as seen with our camera measurements.  We can measure the total number of atoms using the MOT fluorescence with our camera or a photomultiplier tube (PMT); both measurements typically agree within $5\%$ of one another despite using different collection optics, different viewpoints, and having been independently calibrated.  Release and recapture measurements \cite{Chu:1985} taken with the PMT indicate that the MOT atoms follow a cumulative distribution function consistent with a Maxwell-Boltzmann (MB) speed distribution.

Because the type I MOT has a stronger cycling transition strength, it forms a denser and colder MOT, with typical measured densities $n_\mathrm{MOT} \sim 10^{10}~\mathrm{cm}^{-3}$, $1/e$ density radius $r_a \approx 0.025$ cm and temperature $T_\mathrm{MOT} = 0.5 \pm 0.1$ mK.  The type II MOT is larger and warmer, typically having $n_\mathrm{MOT} \sim 10^{9}~\mathrm{cm}^{-3}$, $r_a \approx 0.075$ cm and $T_\mathrm{MOT} = 2.0 \pm 0.5$ mK.  For the results presented here, the type I MOT has a $f_e \approx 33\%$ excited-state population \cite{Shah:2007} and the type II MOT has $f_e \approx 23\%$.

We have established the excited state population using a two-level model-dependent measurement of the effective saturation intensity of the \ce{Na} MOT \cite{Dinneen:1992}.  We are currently experimenting with a hybrid trap model-independent measurement of $f_e$ and plan to publish our findings in the near future.
\vspace{-1em}
\subsubsection{Linear Paul trap}
\label{sec:LPT}
\vspace{-1em}
The ion-trapping part of the hybrid trap consists of a segmented LPT \cite{Drewsen:2000}.  The rf driving field, with amplitude $V_\mathrm{rf} = 80 \pm 2$ V, is applied to the center electrode segments creating a sinusoidally oscillating quadrupole saddle potential.  Oscillating the quadrupole potential at a frequency $\Omega/2\pi = 720$ kHz creates a pseudo-harmonic potential which provides trapping in the transverse dimension \cite{Paul:1990}. The axial (long dimension) confinement is established by a static voltage $V_\mathrm{end} = 30.0 \pm 0.2$ V applied to the end segments.

The evolution of a single trapped ion in an LPT is described by the stable solutions to the Mathieu equation \cite{Champenois:2009}. Each trapped ion undergoes a superposition of fast motion at the driving field's rf angular frequency $\Omega$ and a slow secular motion at an angular frequency $\omega_r$ in the transverse dimensions and $\omega_a$ in the axial dimension \cite{Berkeland:1998,Drewsen:2000,Drakoudis:2006}.

The ion trap is loaded by photoionizing (PI)  an excited \ce{Na}($3\mathrm{P}$) MOT atom with a 405 nm photon.  The PI laser beam has a  $r_{1/e} = 0.20 \pm 0.01$ cm collimated intensity radius and is collinear to one MOT beam.  Therefore, the region of PI is always larger than the MOT, even when the MOT is translated off-axis from the beam as much as $\approx 0.125$ cm. While some background excited \ce{Na} atoms are also PI loaded into the LPT, approximately all of the ions are created directly from the MOT since the MOT density is $\approx 3$ orders of magnitude larger than that of the background \ce{Na} vapor.

The equilibrium temperature of the trapped ion cloud $T_I$, loaded from either the type I or the type II MOT, was determined using \textsc{simion} 7.0 simulations \cite{Manura:2007,Appelhans:2002} of ion clouds containing an ion population $N_I$ of up to $1000$ interacting ions. It takes approximately 1.4 ms or 238 secular periods for the ion cloud to equilibrate. For a detailed discussion of our group's simulations, see Refs.~\cite{Goodman:2012, Sivarajah:2013}.  The most important factor in predicting the ion cloud's thermalized mean secular energy (from which one can assign a temperature, assuming a MB speed distribution) is the size of the MOT when the LPT is loaded via PI from a MOT \cite{Goodman:2012, RaviAPB:2012}.  Because the initial speed of the ions created from the MOT is so small, the total initial energy of the ion cloud is primarily determined by the potential energy of the ions, which is directly related to the initial size of the MOT.  Therefore, since we can accurately measure the size of the \ce{Na} MOT, we can accurately initialize our simulations. The simulation determined the thermalized temperature of the ion cloud loaded from the type I and II MOTs to be $T_I = 140 \pm 10$ K and $T_I = 1070 \pm 30$ K, respectively.  The uncertainty in $T_I$ is only based on the precision of camera measurements of the MOTs' dimensions.
 \begin{figure}[t]
   \centering
   \includegraphics[width=3.25 in]{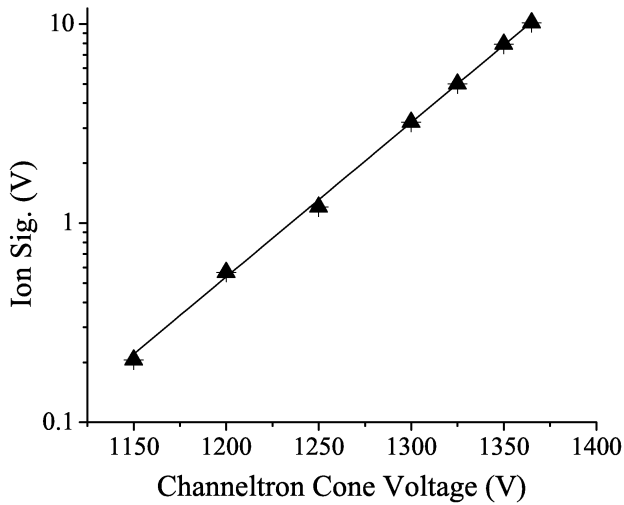}
   \caption{(Color online). We tested for deviation from the expected exponential behavior of the CEM gain as a function of the voltage applied to the CEM detection cone $V_c$ at a fixed ion input current.  We find that the output ion signal (which is proportional to the gain) looks very linear when plotted with a log-linear plot scaling, indicating that we are not saturating the CEM. The uncertainty in the power supply voltage $V_c$ and the ion signal is approximately the size of the plot markers. In the experiments presented here we operate at a $V_c = 1250$ V.}
   \label{fig:CTGain}
\vspace{-0.75em}
\end{figure} 

An undesirable complication with a \ce{Na} MOT hybrid trap is that the MOT continuously forms \ce{Na2+} molecular ions via photoassociative ionization and energetic ($\sim0.5~\mathrm{eV}$) atomic \ce{Na+} is subsequently created via 589 nm photodissociation \cite{Gould:1988,Julienne:1991,Trachy:2007,Tapalian:1994}. To remove the undesired \ce{Na2+} ions, we add to the driving rf voltage a small mass selective resonant quenching (MSRQ) \cite{Hashimoto:2006, Baba:1996, Drakoudis:2006, Sivarajah:2013} voltage with amplitude $V_\mathrm{rad} = 0.625 \pm 0.005$ V  at a frequency $157 \pm 1$ kHz, which corresponds to the measured second harmonic secular frequency $2 \left[ \omega_r/(2\pi) \right]$ for \ce{Na2+}.  The MSRQ signal resonantly drives the secular motion of the co-trapped \ce{Na2+} until the molecular ion's energy exceeds the LPT's trap depth.  As a result, the added MSRQ field continuously quenches the \ce{Na2+} population with little to no off-resonant heating of the trapped \ce{Na+} \cite{Sivarajah:2013}.

Because the \ce{Na+} ions have a closed electronic configuration optical measurements are not possible, so we must destructively measure the trapped ion population. We apply a dipole field to the end segments, which extracts the ion cloud axially out of the trap and into a Channeltron electron multiplier (CEM).  The ion extraction trajectories are controlled by the ion optics, which are determined by the end segment and mesh electrode voltages. The CEM signal goes through a charge-sensitive preamplifier, which produces a signal whose peak voltage is proportional to the number of detected ions.  We will refer to the peak preamplifier voltage as the ``CEM measured ion signal."  Details regarding the calibration of the CEM will be discussed in Sec.~\ref{sec:IonMeas}.

We tested the linearity of the dynamic range of the CEM to ensure that the it was not saturated when detecting large ion populations, $N_I \sim 10^5$. CEMs typically produce linear output in an analog mode when the output current $I_o< 20\%$ of the bias current. The bias current is linearly proportional to the applied CEM cone voltage $V_c$ \cite{Channeltron:1}. The output current $I_o=G I_i$ depends on the gain $G$ and the input current $I_i$, where the gain increases exponentially with increasing $V_c$.

We use a \textsc{megaspiraltron} from Photonis, which has a particularly large bias current, $\approx 160 ~\mathrm{\mu A}$ at $V_c = 2500$ V. By operating at a low $V_c =1250 \pm 6$ V, we reduced the gain exponentially while the bias current falls off linearly, which helps to keep $I_o$ below the $20\%$ limit for large ion signals.

When the CEM saturates, the gain will cease to increase exponentially with increasing $V_c$ for a fixed input ion current $I_i$. We measured the output ion signal, which is $\propto G$ for a fixed $I_i$, as a function of $V_c$ for many different PI intensities, for both MOTs, and at several different ion optic settings.  In all cases we found that for a fixed $I_i$ the log of the ion signal was linear with $V_c$ within the experimental uncertainty, which suggests that the CEM is not saturating (as seen in Fig.~\ref{fig:CTGain}).

We were surprised by this result, because if the ion output current is estimated using the expected order of magnitude for the gain, it is $>20\%$ of the bias current and should be in the nonlinear region. We speculate that not all of the ions are reaching the CEM detection cone.  This sub-unity collection efficiency is not a problem as long as we are approximately losing the same fraction of the total number of ions trapped during each extraction, which we found to be consistent with our data.

We assume the trapped ions adhere to a prolate Gaussian spatial distribution with $1/e$ radii $r_{I,1} = r_{I,2}$ in the transverse dimensions and $r_{I,3} > r_{I,1}$ in the axial dimension. The number of ions loaded into the LPT and the size of the ion cloud will be discussed later in Secs.~\ref{sec:IonMeas} and \ref{sec:CloudSize}, respectively.
%%%%%%%%%%%%%%%%%%%%%%%%%%%%%%%%%%%%
\vspace{-1em}
\subsection{Semiclassical scattering model}
\label{sec:ScatModel}
\vspace{-1em}
Fully quantal \textit{ab intio} calculations have shown that both elastic and charge-exchange two-body ion-neutral scattering cross sections follow semiclassical power-law functions of energy in the $10^4$ - $10^{-6}$ K temperature range. Two-body charge-exchange occurs when an electron is transferred from the ion to the atom. The charge-exchange process is considered resonant if the internal states are exchanged without changing the total internal energy.  A collision is elastic if no charge exchange occurs and the total kinetic energy is conserved.

The total scattering cross section $\sigma_\mathrm{tot}$ has a
\begin{equation}
\sigma_\mathrm{tot} = \sigma_\mathrm{el}+\sigma_\mathrm{ce} = C_\mathrm{tot} E^{-1/3}
\label{eq:sigmatot}
\end{equation}
relative collision energy $E$ dependence, where $C_\mathrm{tot}$ is the total scattering proportionality constant, $\sigma_\mathrm{el}$ is the elastic scattering cross section, and $\sigma_\mathrm{ce}$ is charge-exchange scattering cross section \cite{Zhang:2009}.  The total scattering constant $C_\mathrm{tot}$ is proportional to $(\mu {C_4}^2)^{1/3}$, where $\mu$ is the reduced mass of the two-body collision.  Equation (\ref{eq:sigmatot}) is incorrectly stated as only the elastic scattering cross sections in Refs.~\cite{Cote:2000,Makarov:2003}, but is correctly identified as the total cross section in Ref.~\cite{Zhang:2009}. However, the distinction makes little difference in our case, because $\sigma_\mathrm{ce} \ll \sigma_\mathrm{el}$, as found in Ref.~\cite{Cote:2000}.

By averaging over the relative energy distribution, the total rate coefficient for the ion-neutral collisions $k_\mathrm{ia}$ can be expressed in atomic units as
\begin{align}
k_\mathrm{ia} &= \left < \sigma_\mathrm{tot} \sqrt{\frac{2E}{\mu}}\right >_E \nonumber \\
                        \nonumber \\
                        &= C_\mathrm{tot} \Gamma \left ( \frac{5}{3} \right )\sqrt{\frac{8}{\pi \mu}} \left ( k_B T_I \right )^{1/6},
\label{eq:kia}
\end{align}
where $\Gamma$ is the gamma function, and $k_B$ is the Boltzmann constant \cite{Makarov:2003}.  We make the standard assumption that the ions have a MB speed distribution \cite{Ravi:2012, Lee:2013, Chen:2007} and that because $T_\mathrm{MOT} \ll T_I$, the relative speed distribution is approximated well by the ion cloud's speed distribution \cite{Smith:2014, Lee:2013, Haze:2013}.  According to Eq.~(\ref{eq:kia}), $k_\mathrm{ia}$  depends weakly on $T_I$. Therefore, determining $T_I$ via \textsc{simion} simulations is sufficiently accurate for determining the theoretical total rate coefficient to be used for comparison with experiment.

The values for the total collision rate coefficient for the \ce{Na+}-- \ce{Na} system can be found in Table~\ref{tab:kia}, where the input for  Eq.~(\ref{eq:kia}) comes from the power-law fits found in Ref.~\cite{Cote:2000} at the relevant temperatures $T_I$ associated with our experiment.  We use the scaling of $C_4$ to determine the excited \ce{Na}(3$\mathrm{P}$) $C_\mathrm{tot}$ value.
\vspace{-0.5em}
\begin{table}[h]
\caption{Table of semiclassically predicted total (elastic and charge exchange) ion-atom scattering rate coefficients $k_\mathrm{ia}$ for both excited ($3\mathrm{P}$) and ground state ($3\mathrm{S}$) \ce{Na} on ground state \ce{Na+} at the experimentally relevant ion cloud temperatures $T_I$.  The total rate coefficient is calculated using Eq.~(\ref{eq:kia}). The uncertainty in $k_\mathrm{ia}$ is due to the propagated uncertainty in $T_I$.}
\centering
\begin{ruledtabular}
\begin{tabular}{c|c|c|c|c}
Species & $C_4$ (a.u.) & $C_\mathrm{tot}$ (a.u.) & $T_I$ (K) & $k_\mathrm{ia} \left ( \mathrm{cm}^3/\mathrm{s} \right )$ \\
\hline
\multirow{2}{*}{\ce{Na}(3$\mathrm{S}$)-\ce{Na+}} & \multirow{2}{*}{162.7\footnotemark[1]} & \multirow{2}{*}{4174\footnotemark[1]} & 140(10) & $7.00(8) \times 10^{-8}$ \\
&&& 1070(30) &  $9.82(5) \times 10^{-8}$ \\
\hline
\multirow{2}{*}{\ce{Na}(3$\mathrm{P}$)-\ce{Na+}} & \multirow{2}{*}{361.4\footnotemark[2]} & \multirow{2}{*}{7106} & 140(10) & $1.19(1) \times 10^{-7}$ \\
&&&1070(30) &  $1.67(1) \times 10^{-7}$ \\
\end{tabular}
\end{ruledtabular}
\label{tab:kia}
\footnotetext[1]{Reference \cite{Cote:2000}}
\footnotetext[2]{Reference  \cite{Mitroy:2010}}
\end{table}
%%%%%%%%%%%%%%%%%%%%%%%%%%%%%%%%%%%%
\vspace{-2em}
\subsection{Experimental Scattering Model}
\label{sec:ExpMod}
\vspace{-1em}
When the \ce{Na} MOT is overlapped with the ion cloud in the hybrid trap, \ce{Na+}-- \ce{Na} elastic and resonant non-radiative charge-exchange collisions will occur within the volume of overlap. Because the trap depth of the MOT is fairly small $\sim 0.1$ K \cite{Raab:1987} we can make the approximation that every elastic or charge-exchange collision will result in the loss of a MOT atom \cite{Lee:2013}.

We can check the validity of this approximation with a simple calculation. Let us consider two-body hard sphere collisions \cite{He:1997} between $T \approx 0$ K (near delta function speed distribution) \ce{Na} atoms held within a 0.1 K deep MOT and a $T_I = 500$ K \ce{Na+} ion cloud.  For charge-exchange collisions we need only integrate over the \ce{Na+} ion cloud MB speed distribution from the MOT trap depth to infinity. We find that more than $99\%$ of the ion population has a velocity large enough to cause an atom to be lost from the trap after a charge-exchange collision.

For two-body hard sphere elastic scattering, we can assume an isotropic solid angle center-of-mass scattering angle distribution. Again, by integrating over the entire angular distribution and the relevant speeds of the MB speed distribution we find that on average more than $99\%$ of the ion population will eject a MOT atom during an elastic ion-atom collision. The only consequence of assuming that ion-atom collisions cause MOT loss with unit efficiency is that the experimentally determined value for $k_\mathrm{ia}$ will be systematically underestimated, but our simple calculations suggest this systematic error should be negligible.

We define the loss rate per atom from the MOT due to ion-atom collisions $\gamma_\mathrm{ia}$ to be
\begin{align*}
\gamma_\mathrm{ia} & = k_\mathrm{ia} \left < n \right > \\
\\
                                   &= k_\mathrm{ia} N_I \prod_{i=1}^3 \int_{-\infty}^\infty \left ( \frac{e^{-(x_i - x_{0,i})^2/r_a^2}}{r_a \sqrt{\pi}} \right ) \left ( \frac{e^{-x_i^2/r_{I,i}^2}}{r_{I,i} \sqrt{\pi}} \right ) dx_i
\end{align*} 
where $\left < n \right >$ is the average ion density experienced by the MOT \cite{Grier:2009}, $x_i$ is the distance from the center of the ion cloud in the $i=1, 2,~\mathrm{or}~3$ dimension, and $x_{0,i}$ is the center position of the MOT relative to the center of the ion cloud in the $i^{th}$ dimension.  Upon integrating over the ion and atom cloud Gaussian spatial distributions, we arrive at 
\begin{equation}
\gamma_\mathrm{ia}=\frac{ k_\mathrm{ia} N_I C}{V_\mathrm{ia}},
\label{eq:gia}
\end{equation}  
which shows that the loss rate is proportional to the total trapped number of ions $N_I$, 
the relative concentricity function
\begin{equation}
C = e^{- \left (x_{0,1}^2+x_{0,2}^2 \right )/\left ( r_a^2 + r_{I,1}^2 \right )} e^{- x_{0,3}^2/ \left ( r_a^2 + r_{I,3}^2 \right )},
\label{eq:C}
\end{equation}
and inversely proportional to the addition in quadrature of the effective volumes of the ion and atom clouds
\begin{equation}
V_\mathrm{ia} = \pi^{3/2} \left ( r_a^2+r_{I,1}^2 \right ) \sqrt{r_a^2+r_{I,3}^2}.
\label{eq:Via}
\end{equation}
Equation \ref{eq:C} is equal to unity when the MOT is perfectly centered on the ion cloud.  If we approximate $r_a \ll r_I$, we arrive at the same expression for $\gamma_\mathrm{ia}$ used in Ref.~\cite{Lee:2013}. We can experimentally measure the loss rate $\gamma_\mathrm{ia}$, the number of ions $N_I$, and the volumes that make up $V_\mathrm{ia}$, which gives enough information to solve for $k_\mathrm{ia}$ using Eq.~(\ref{eq:gia}).

We followed Ref.~\cite{Lee:2013}'s choice to measure the loss rate $\gamma_\mathrm{ia}$ when the LPT is saturated, which has three advantages.  First, the saturated ion cloud volume $\tilde{V}_I$ remains constant for each measurement, thereby making the saturated addition in quadrature of the ion and atom cloud volumes $\tilde{V}_\mathrm{ia}$ time independent.   Second, because the LPT is in steady-state, the ion population $\tilde{N}_I$ can be approximated as time independent. Third, the saturated LPT holds the largest possible number of ions $\tilde{N}_I$ (for a given cloud temperature $T_I$ and trap settings). Therefore, the saturated LPT maximizes $\gamma_\mathrm{ia}$, which gives the greatest experimental resolution of $k_\mathrm{ia}$.  
%%%%%%%%%%%%%%%%%%%%%%%%%%%%%%%%%%%%
%%%%%%%%%%%%%%%%%%%%%%%%%%%%%%%%%%%%
\vspace{-1em}
\section{EXPERIMENT AND  RESULTS}
\label{sec:ExpResults}
\vspace{-1em}
\subsection{\ce{Na} MOT loading measurements}
\label{sec:MOTMeas}
\vspace{-1em}
In the temperature-limited regime \cite{Townsend:1995, Wipple:2001}, the volume of the MOT $V_\mathrm{MOT}$ remains constant while the MOT density $n_\mathrm{MOT}$ increases linearly with atom population $N_a$. Collisions between two MOT atoms lead to a non-exponential two-body loss rate $\beta n_\mathrm{MOT}$ \cite{Prentiss:1988}, while collisions with constant density background \ce{Na} atoms result in a linear loss rate $\gamma_b$.  Because we operate in the temperature-limited regime, we model the MOT loading behavior with a non-linear rate equation
\begin{equation}
\frac{dN_a}{dt} = L_\mathrm{MOT} - \gamma_t N_a - \frac{\beta}{V_\mathrm{MOT}} N_a^2,
\label{eq:dtMOT}
\end{equation}
where $ L_\mathrm{MOT}$ is the constant rate at which atoms are loaded into the MOT and $\gamma_t$ is the total single-body linear loss rate \cite{Wipple:2001}.  The solution to Eq.~(\ref{eq:dtMOT}) is
\begin{equation}
N_a(t)=\frac{2L_\mathrm{MOT} \left ( 1-e^{-\gamma_e t} \right )}{\gamma_e + \gamma_t + \left ( \gamma_e - \gamma_t \right ) e^{-\gamma_e t}},
\label{eq:MOTsol}
\end{equation}
where 
\begin{equation}
\gamma_e = \sqrt{\gamma_t^2+\frac{4 \beta L_\mathrm{MOT}}{V_\mathrm{MOT}}}.
\label{eq:ge}
\end{equation}

We found that using Eq.~(\ref{eq:MOTsol}) significantly improved our fits to the MOT fluorescence loading data, as opposed to the more commonly used linear rate equation \cite{Duncan:2001, Wipple:2001, Lee:2013}. However, to reduce the number of free parameters, we found that constraining $\beta$ to a value of $\approx 1.0 \times 10^{-11}$ cm$^3$/s for the type I MOT and a value of $\approx 1.0 \times 10^{-10}$ cm$^3$/s for the type II MOT gave the most consistent fits.  These values are fairly close to the previously reported value of $\beta$ for a \ce{Na} MOT of $4 \times 10^{-11}$ cm$^3$/s, which has a factor of five uncertainty \cite{Prentiss:1988}.

\begin{figure}[t]
   \centering
   \includegraphics[width=3.25 in]{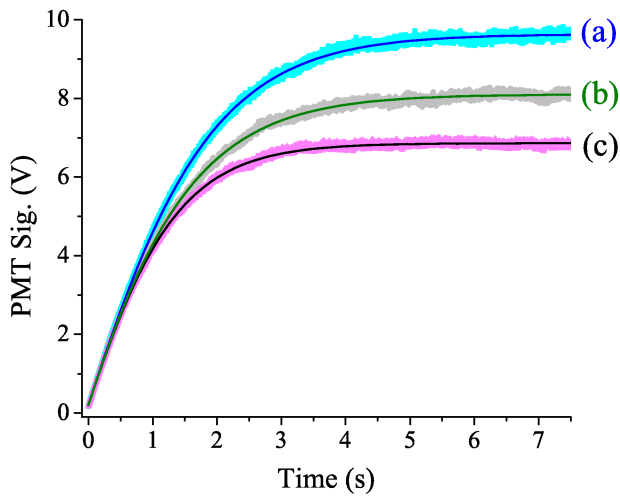}
   \caption{(Color online). Fluorescence from a type II MOT as it loads, with the corresponding fits to Eq~(\ref{eq:MOTsol}).  Curve (a) shows the raw PMT data (light blue) and fit (royal blue) of an isolated MOT loaded with a total loss rate $\gamma_t=\gamma_b$. Curve (b) shows the raw PMT data (gray) and fit (green) of a MOT loaded with PI ($I_\mathrm{pi} \approx$ 80 mW/cm$^2$), making $\gamma_t=\gamma_b+\gamma_\mathrm{pi}$. Curve (c) shows the raw PMT data (magenta) and fit (black) of a MOT loaded with the same PI intensity as curve (b), but the MOT is also immersed in a saturated LPT ion cloud, making $\gamma_t = \gamma_b +\gamma_\mathrm{pi}+ \gamma_\mathrm{ia}$.}
   \label{fig:PMT}
\vspace{-0.75em}
\end{figure}

We found the MOT loading rate $L_\mathrm{MOT}$ to be insensitive to the presence of PI or an ion cloud.  Similar behavior has been observed elsewhere \cite{Lee:2013, Duncan:2001}.  Furthermore, an experiment that modeled changes to $L_\mathrm{MOT}$ in a \ce{Na} MOT due to PI found that the modification was small \cite{Wipple:2001}, therefore we neglect it in the interest of simplicity.

Figure~\ref{fig:PMT} shows the \ce{Na} fluorescence measured by the PMT and fit with Eq.~(\ref{eq:MOTsol}) when the type II MOT is loaded at three different loss rates $\gamma_t$. The type I MOT loading curves are qualitatively identical to that of the type II MOT. The total loss rate depends upon the loss mechanisms that are present at the time the MOT is loaded.   Figure \ref{fig:PMT} curve (a) is for an isolated MOT loaded from background \ce{Na} vapor $\gamma_t = \gamma_b$.

When the MOT is also experiencing PI, there is an additional loss rate $\gamma_\mathrm{pi}$, which increases the total loss rate $\gamma_t = \gamma_b + \gamma_\mathrm{pi}$, as is the case in Fig.~\ref{fig:PMT} curve (b).  At low enough PI intensity $I_\mathrm{pi}$, the PI loss rate $\gamma_\mathrm{pi}$ is linearly proportional to $I_\mathrm{pi}$ and can be expressed as
\begin{equation}
\gamma_\mathrm{pi} = \frac{\sigma_\mathrm{pi} f_e I_\mathrm{pi}}{h \nu_\mathrm{pi}} = \zeta I_\mathrm{pi},
\label{eq:gpi}
\end{equation}
where $\sigma_\mathrm{pi}$ is the PI cross section, $h$ is Plank's constant, and $\nu_\mathrm{pi}$ is the frequency of the PI radiation, and again $f_e$ is the fraction of MOT atoms in the excited state \cite{Wipple:2001, Lee:2013,Duncan:2001}.

\begin{figure}[t]
   \centering
   \includegraphics[width=3.25 in]{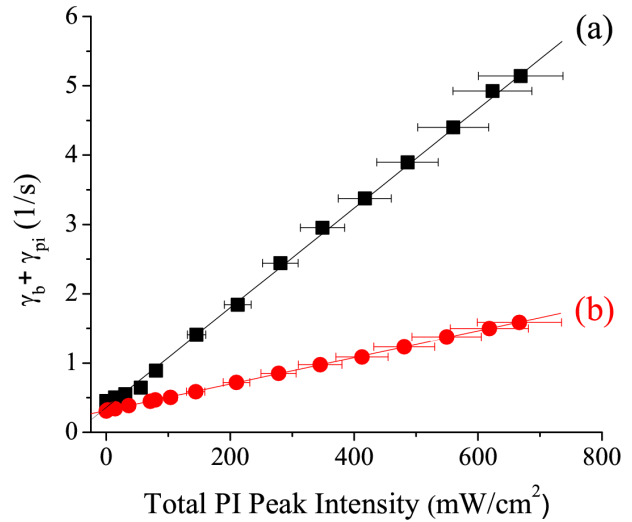}
   \caption{(Color online). Plot of the total MOT loss rate in the presence of PI ($\gamma_t = \gamma_b+\gamma_\mathrm{pi}$) as a function of the total peak PI intensity and corresponding linear fits.  Curve (a) shows type I MOT data and curve (b) shows type II MOT data.  The y-intercepts (at $I_\mathrm{pi} = 0$) are $\gamma_b$ and the slopes are proportional to $\sigma_\mathrm{pi} f_e$.  The statistical uncertainty in the rates are smaller than the plot markers.  The uncertainty in the intensity is primarily due to power fluctuations and the precision of the beam waist measurement.}
   \label{fig:pi}
\vspace{-0.75em}
\end{figure}

We find that $\gamma_\mathrm{pi}$ is linear with $I_\mathrm{pi}$ for both MOTs over the full PI intensity range achieved with our setup, as shown in Fig.~\ref{fig:pi}.  The y-intercept (at $I_\mathrm{pi} = 0$) of Fig.~\ref{fig:pi} is equal to $\gamma_b$, while the slope can be used to determine $\sigma_\mathrm{pi}$.  The slope of curve (a) for the type I MOT gives $\sigma_\mathrm{pi} = 1.1 \pm 0.2 \times 10^{-17}$ cm$^2$ and the slope of curve (b) for the type II MOT gives  $\sigma_\mathrm{pi} = 4.1 \pm 0.9 \times 10^{-18}$ cm$^2$.  Both results are fairly close to the previously reported experimental value of $\sigma_\mathrm{pi} = 9.1 \pm 1.4 \times 10^{-18}$ cm$^2$ for 404 nm PI radiation from Ref.~\cite{Wipple:2001}, which also had very good agreement with theory \cite{Preses:1985}.

The final loss mechanism is from ion-atom collisions between the MOT and the saturated LPT ion cloud, as seen in Fig.~\ref{fig:PMT} curve (c). These collisions introduce an additional term, which increases the loss rate to $\gamma_t = \gamma_b +\gamma_\mathrm{pi} + \gamma_\mathrm{ia}$.  The ion-atom loss rate $\gamma_\mathrm{ia}$ at each PI intensity $I_\mathrm{pi}$ was determined by subtracting the loss rate measured with  the PI laser on and the LPT turned off from measurements with both the PI laser and the LPT turned on.

Unlike the experimental sequence presented in Ref.~\cite{Lee:2013}, before taking the MOT loading data in Fig.~\ref{fig:PMT} curve (c), the LPT is pre-loaded from the MOT until the LPT is saturated.  The MOT is then briefly unloaded by blocking one of the retro-reflected 589 nm beams with an electronic shutter.  Last, the MOT is reloaded while immersed in the saturated ion cloud. The PI laser remains on during the entire sequence to ensure the LPT remains saturated.  By pre-loading the LPT to saturation before taking the PMT measurement, we can approximate the ion cloud surrounding the MOT as having a constant volume in each measurement $\tilde{V}_I$. We can also approximate the density $\tilde{N}_I/\tilde{V}_I$ as time independent during a loading measurement, making $\gamma_\mathrm{ia}$ time independent.

\begin{figure}[t]
   \centering
   \includegraphics[width=3.25in]{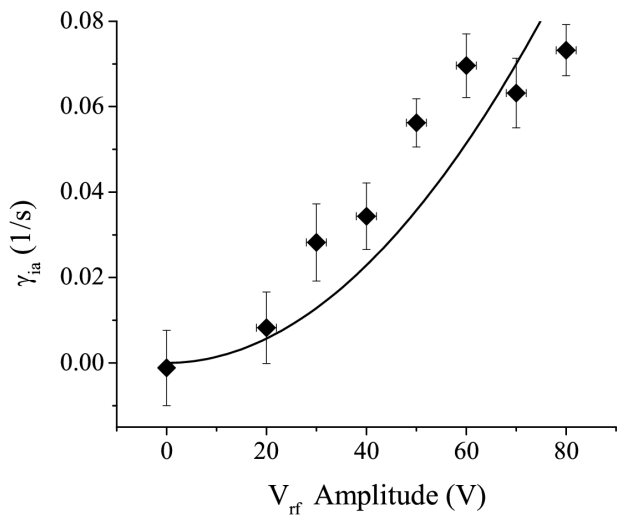}
   \caption{(Color online). Plot of the ion-atom loss rate $\gamma_\mathrm{ia}$ as a function of increasing rf voltage amplitude $V_\mathrm{rf}$.  Because the saturated LPT density increases quadratically with $V_\mathrm{rf}$,  the ion-atom loss rate also appears to increase quadratically. Although, the data could also be interpreted as being linearly propostional to $V_\mathrm{rf}$.}
   \label{fig:giaVrf}
\vspace{-1.1em}
\end{figure}

To achieve the greatest experimental resolution for $\gamma_\mathrm{ia}$, we worked at a high rf voltage amplitude $V_\mathrm{rf}$ that puts us close to the edge of the Mathieu equation's stability region \cite{Drewsen:2000}.  We found that $\gamma_\mathrm{ia}$ increased approximately quadratically with $V_\mathrm{rf}$, as suggested in Fig.~\ref{fig:giaVrf}. We can rationalize the proportionality between $\gamma_\mathrm{ia}$ and $V_\mathrm{rf}$ through the following scaling arguments.

When the LPT is saturated, we can determine the size of the ion cloud by equating the effective LPT trap depth to the energy of the outermost ion in a simple harmonic potential with a frequency equal to the secular frequency \cite{Smith:2014,Lee:2013}.  For an idealized single particle in a perfect quadrupole field, the LPT trap depth is proportional to $(V_\mathrm{rf})^2$ \cite{Raizens:1992}, as is the square of the secular frequency.  Therefore, we expect the saturated size of the cloud $\tilde{V}_I $ to be insensitive to $V_\mathrm{rf}$.  By equating the LPT's spring force to the ion cloud's Coulomb repulsion (for an infinite cylinder of charge), it can be shown that the saturated number of trapped ions $\tilde{N}_I \propto (V_\mathrm{rf})^2$.  Therefore, since $\gamma_\mathrm{ia} \propto \tilde{N}_I/ \tilde{V}_I \implies \gamma_\mathrm{ia} \propto (V_\mathrm{rf})^2$.

\begin{figure}[t]
   \centering
   \includegraphics[width=3.25in]{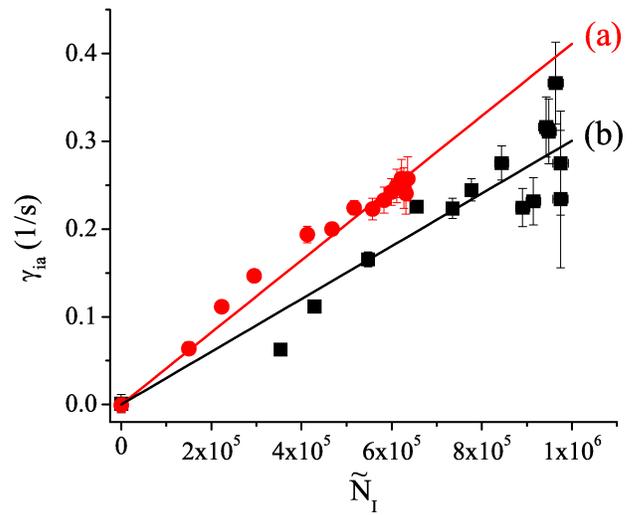}
   \caption{(Color online). Plot of the ion-atom loss rate as a function of the saturated LPT steady-state ion population and corresponding linear fit to Eq.~(\ref{eq:gia}).  Curve (a) shows type II MOT data and curve (b) shows type I MOT data.  The type II MOT has larger loss rates and fewer steady-state ions at saturation because it produces a hotter lower density ion cloud that has a slightly larger $k_\mathrm{ia}$. The uncertainty in the measurements are discussed in the text.}
   \label{fig:giaNI}
\vspace{-0.75em}
\end{figure}

As seen in Fig.~\ref{fig:giaNI}, we plot $\gamma_\mathrm{ia}$ as a function of the steady-state LPT ion population $\tilde{N}_I$. As predicted by Eq.~(\ref{eq:gia}), these quantities are linearly proportional. The observed linearity supports our assumption from Sec.~\ref{sec:LPT} that the fraction of extracted ions that miss the CEM is fairly constant, if we assume Eq.~(\ref{eq:gia}) to be correct. 

The fractional uncertainty in the measurement of $\gamma_\mathrm{ia}$ appears to increase with $I_\mathrm{pi}$ (Fig.~\ref{fig:giaIp}) or steady-state ion population (Fig.~\ref{fig:giaNI}).  This can be explained by the fact that $\gamma_\mathrm{ia}$ is the difference of two measurements whose individual fractional uncertainty remains fairly constant.  However, since the difference between these measurements saturates, as seen in Fig.~\ref{fig:giaIp}, the fractional uncertainty in the difference $\gamma_\mathrm{ia}$ must increase.

We will discuss the LPT loading behavior including how the steady-state ion population seen in Fig.~\ref{fig:giaNI} was determined in Sec.~\ref{sec:IonMeas}.  Finally, using the slopes from Fig.~\ref{fig:giaNI} and the ion cloud size, which we will discuss in Sec.~\ref{sec:CloudSize}, we will have enough information to determine the rate coefficient $k_\mathrm{ia}$.
%%%%%%%%%%%%%%%%%%%%%%%%%%%%%%%%%%%%
\vspace{-1em}
\subsection{LPT \ce{Na+} loading and decay measurements}
\label{sec:IonMeas}
\vspace{-1em}
\subsubsection{LPT loading}
\label{sec:LPTload}
\vspace{-1em}
\begin{figure}[b]
   \centering
   \includegraphics[width=3.25 in]{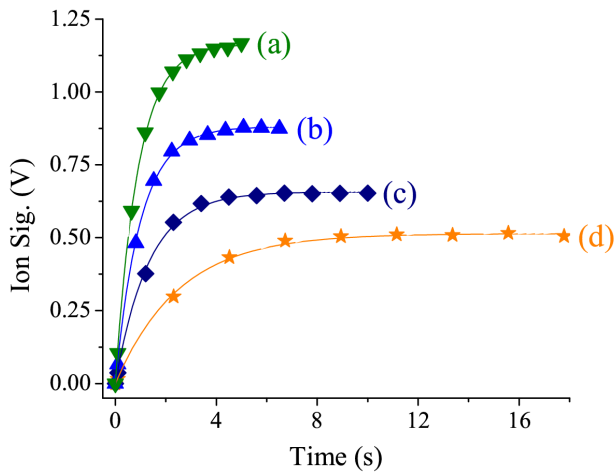}
   \caption{(Color online). CEM measured LPT loading (from the type I MOT) as a function of time and corresponding two-parameter fits to the solution to rate Eq.~(\ref{eq:LI}). Each curve corresponds to a different PI intensity: Curve (a) is measured with $I_\mathrm{pi} \approx 670$ mW/cm$^2$, curve (b) is with $I_\mathrm{pi} \approx 108$ mW/cm$^2$, curve (c) is with $I_\mathrm{pi}\approx 42$ mW/cm$^2$, and curve (d) is with $I_\mathrm{pi} \approx 15$ mW/cm$^2$. The uncertainties are smaller than the size of the plot markers.}
   \label{fig:LoadN1}
\vspace{-0.75em}
\end{figure}
According to Eq.~(\ref{eq:gia}), $\gamma_\mathrm{ia}$'s dependence on $I_\mathrm{pi}$ comes from $\tilde{N}_I$'s dependence on $I_\mathrm{pi}$.  Due to experimental difficulties with CEM saturation, Ref.~\cite{Lee:2013} attempted to derive an LPT loading model that determined $\tilde{N}_I$ solely from MOT fluorescence measurable quantities, such as the MOT atom population $\tilde{N}_a$, the PI MOT loss rate $\gamma_\mathrm{pi}$, and the ion-atom MOT loss rate $\gamma_\mathrm{ia}$.

They modeled the LPT loading with the linear rate equation
\begin{equation}
\frac{dN_I}{dt} = L_I - \lambda N_I,
\label{eq:dtLPT1}
\end{equation}
where $L_I$ is the LPT ion loading rate and $\lambda$ is the LPT ion loss rate. We find good agreement between Ref.~\cite{Lee:2013}'s LPT rate equation [our Eq.~(\ref{eq:dtLPT1})] and our experimental data, as seen in Fig.~\ref{fig:LoadN1}, which shows typical LPT loading curves taken with the CEM at four different $I_\mathrm{pi}$ intensities loaded from the type II MOT.  The fits use $L_I$ and $\lambda$ as free fitting parameters, which makes the steady-state ion population the ratio of the two fitting parameters $\tilde{N}_I = L_I/\lambda$.  Experimentally, for each PI laser intensity we preset the MOT into a steady-state atom population $\tilde{N}_a$ with the PI laser on before turning on the LPT.  The LPT is loaded from the MOT for a fixed time and then the ions are immediately extracted and detected.  This procedure is repeated with increasing loading times until the LPT has reached its steady-state ion population.

Reference~\cite{Lee:2013} argues that the number of MOT atoms lost are proportional to the number of ions gained by the LPT. 
Accordingly, the loss rate lambda is equated to the ion-atom MOT loss rate $\gamma_\mathrm{ia}$ and the loading rate is modeled with a linear dependence on $I_\mathrm{pi}$,
\begin{equation}
L_I=N_a \gamma_\mathrm{pi} = N_a \zeta I_\mathrm{pi},
\label{eq:LILee}
\end{equation}  
which diverges as $I_\mathrm{pi} \rightarrow \infty$.  Because the LPT cannot hold an infinite number of ions, they introduce an intensity loss coefficient $\kappa$ and the PI intensity differential equation
\begin{equation}
\frac{dN_I}{dI_\mathrm{pi}} = \frac{N_a\zeta}{\gamma_\mathrm{ia}} \left ( 1 - e^{-\gamma_\mathrm{ia}t} \right )- \kappa N_I.
\label{eq:NILee}
\end{equation}
In deriving Eq.~\ref{eq:NILee} and its solution (as $t \rightarrow \infty$)
\begin{equation}
\tilde{N}_I = \frac{\tilde{N}_a\zeta}{\gamma_\mathrm{ia}\kappa} \left ( 1 - e^{-\kappa I_\mathrm{pi}} \right ),
\label{eq:NIssLee}
\end{equation}
 Ref.~\cite{Lee:2013} appears to make the approximation that $dN_a/dI_\mathrm{pi} = d\gamma_\mathrm{ia}/dI_\mathrm{pi} \approx 0$.

Because the MOT is much smaller than the trapping volume of the LPT, every PI ion created from the MOT can be considered loaded into the LPT. However, unlike Ref.~\cite{Lee:2013}, we consider PI intensity dependence of $N_a$ according to Eq.~(\ref{eq:MOTsol}), and we do not make the assumption that $dN_a/dI_\mathrm{pi} \approx 0$. Also, because we allow the MOT to come to steady-state $\tilde{N}_a$ before turning on the LPT, our ion trap loading rate is
\begin{align}
L_I &= \tilde{N}_a \gamma_\mathrm{pi} \nonumber \\
 &\approx  \frac{2L_\mathrm{MOT} \zeta I_\mathrm{pi}}{\gamma_b+\zeta I_\mathrm{pi} + \sqrt{\left ( \gamma_b+\zeta I_\mathrm{pi} \right )^2 + \frac{4 \beta L_\mathrm{MOT}}{ V_\mathrm{MOT}}}}.
\label{eq:LI}
\end{align}
By including the atom number's PI intensity dependence we see that the ion trap loading rate already saturates as $I_\mathrm{pi} \rightarrow \infty$ without the need for introducing an intensity-loss coefficient $\kappa$.
\begin{figure}[t]
   \centering
   \includegraphics[width=3.25 in]{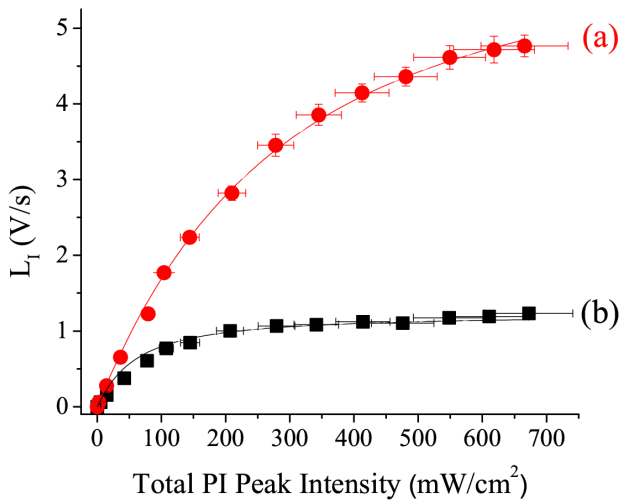}
   \caption{(Color online). CEM measured LPT loading rate as a function of $I_\mathrm{pi}$ and corresponding fit to Eq.~(\ref{eq:LI}).  Curve(a) shows the LPT loaded from the type II MOT and curve (b) shows the LPT loaded from the type I MOT.  The fit to Eq.~(\ref{eq:LI}) is only a single parameter y-scaling constant, whose value is the CEM calibration.  All other parameters are independently determined from MOT fluorescence measurements.}
   \label{fig:LI}
\vspace{-0.75em}
\end{figure}

Figure \ref{fig:LI} shows the ion trap loading rate measured with the CEM as a fucntion of $I_\mathrm{pi}$, when loaded from both the type I and II MOTs. We see that $L_I$ is not linearly proportional to $I_\mathrm{pi}$, as Eq.~(\ref{eq:LILee}) would suggest. We have fit $L_I$ to Eq.~(\ref{eq:LI}), with only a single fitting parameter to scale the y-axis.  All other parameters are independently determined from the MOT fluorescence measurements discussed in Sec.~\ref{sec:MOTMeas}. The single parameter y-scaling fit result gives the CEM calibration.  The type I MOT [curve (b)] has a calibration result of $1.19 \pm 0.02 \times 10^{-6}$ V/ion and the type II MOT [curve (a)] gives $2.70 \pm 0.01 \times 10^{-6}$ V/ion.  The calibrations are fairly close.  We used the calibration for the corresponding MOT when calculating our results.

For simplicity, like Ref.~\cite{Lee:2013}, we ignored $\tilde{N}_a$'s dependence on $\gamma_\mathrm{ia}$, as this is only a small correction, since $\gamma_b + \gamma_{pi} \gg \gamma_\mathrm{ia}$.  By ignoring this term, $L_I$ does not depend on $N_I$, which makes solving Eq.~(\ref{eq:dtLPT1}) much easier.
\begin{figure}[b]
   \centering
   \includegraphics[width=3.25in]{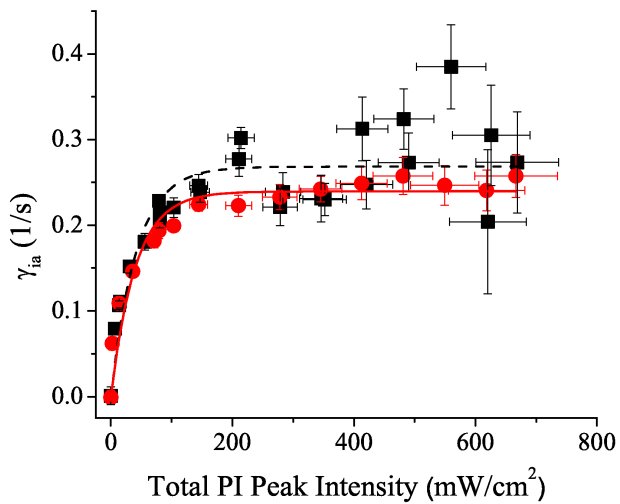}
   \caption{(Color online). Plot of the ion-atom loss rate $\gamma_\mathrm{ia}$ as a function of PI intensity for the type I MOT (black) squares and the type II MOT (red) circles.  We find that $\gamma_\mathrm{ia}$ measurements of the MOT fluorescence are not equivalent to the LPT loss rate measurements seen in Fig.~\ref{fig:lambda}. Furthermore, we show fits to the type I (black dashed line) and type II (solid red line) MOT data obtained as the solution of Ref.~\cite{Lee:2013}'s intensity-loss rate equation.}
   \label{fig:giaIp}
\vspace{-0.75em}
\end{figure}
\begin{figure}[b]
   \centering
   \includegraphics[width=3.25 in]{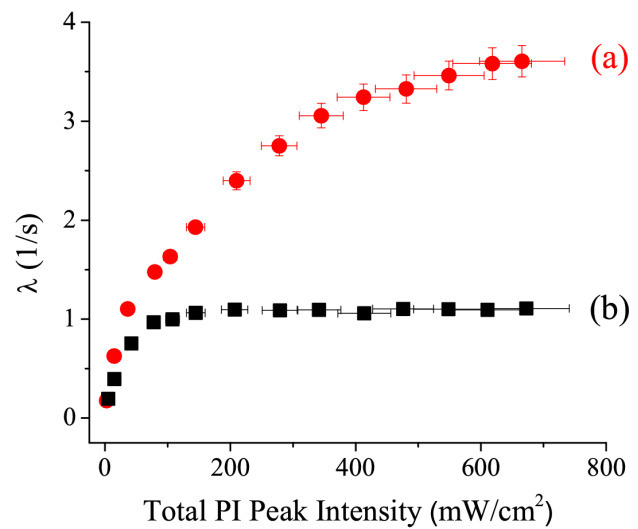}
   \caption{(Color online). Plot of the LPT loss rate $\lambda$ as a function of $I_\mathrm{pi}$.  Each value is determined from fits to loading curves like the ones seen in Fig.~\ref{fig:LoadN1}.  Curve (a) is the $\lambda$ fitting value when the LPT is loaded with the type II MOT and curve (b) is the $\lambda$ fitting value when the LPT is loaded with the type I MOT.}
   \label{fig:lambda}
\vspace{-0.75em}
\end{figure}

We find that the solution to Ref~\cite{Lee:2013}'s intensity-loss coefficient model [our Eq.~(\ref{eq:NIssLee})] for determining the steady-state ion population, which is linearly proportional to $\gamma_\mathrm{ia}$ [according to our Eq.~(\ref{eq:gia})], fits within the experimental error, as shown in Fig.~\ref{fig:giaIp}. We found the agreement to be surprising, since the model's derivation required that $d\gamma_\mathrm{ia}/dI_\mathrm{pi} \approx 0$, which seems inconsistent with the lower PI intensity $\gamma_\mathrm{ia}$ results.  We do find a small systematic difference between our experimental data and the intensity-loss coefficient model. The fits slightly overshoot the data at the knee of the curve and then the fits undershoot the data at the high intensity end of the curve.  The discrepancy is small (as compared with the error bars) but systematic, since it appears in every data run that we have performed for both MOTs. However, it is understandable that this small discrepancy was not observed in Ref.~\cite{Lee:2013}, since the PI intensities used were two orders of magnitude smaller in that study and thus the nearly saturated regime seen in Fig.~\ref{fig:giaIp} was not reached.

As we mentioned before, the LPT loss rate $\lambda$ was equated with $\gamma_\mathrm{ia}$ in Ref.~\cite{Lee:2013}. Unfortunately, we find this to be inconsistent with our data. By comparing Fig.~\ref{fig:giaIp} and Fig.~\ref{fig:lambda}, we see that $\lambda$ does not have the same $I_\mathrm{pi}$ dependence as $\gamma_\mathrm{ia}$. Additionally, $\lambda$ is an order of magnitude larger than $\gamma_\mathrm{ia}$. The ion-atom MOT loss $\gamma_\mathrm{ia}$ goes to zero as PI intensity is decreased.  By equating $\lambda$ with $\gamma_\mathrm{ia}$, Ref.~\cite{Lee:2013} suggests that the trap loss would also go to zero without PI or without the MOT, which is  inconsistent with the fact that the LPT always exhibits some trap loss.

A resonant charge-exchange collision results in a \ce{Na+} with an energy close to that of a MOT atom. Elastic collisions with MOT atoms may cause the ion to gain energy but more often result in a lower energy. Because collisions with neutrals, on average, reduce the energy of an ion, only a very small fraction of these collisions cause an ion to be ejected from the very deep (as compared to the MOT) trapping potential of the LPT. Also, elastic and resonant charge-exchange collisions cause no net increase in the number of trapped ions, so for a saturated LPT ion-atom collisions do not necessarily lead to ion loss. Therefore, it is unlikely that an ion-neutral collision will cause an ion to be ejected suggesting that $\gamma_\mathrm{ia}$ should not be equivalent to $\lambda$.

We suggest that $\lambda$'s apparent PI intensity dependence is actually due to a dependence on $N_I$. Therefore, the LPT steady-state population $I_\mathrm{pi}$ dependence comes entirely from $L_I$. If $\lambda$ depends on $N_I$, this would mean that the LPT loss has a space charge dependence, despite the fact that we are operating in the low coupling regime $\Gamma \ll 1$, where $\Gamma$ is the ratio of the nearest-neighbor Coulomb repulsion to the average thermal energy \cite{Champenois:2009}.

To incorporate the effects of two-body collisions, which to lowest order are proportional to the number of trapped ions, we approximated $\lambda$'s ion number dependence as
\begin{equation}
\lambda \approx \lambda_1 + \lambda_2 N_I,
\label{eq:lambda}
\end{equation}
where $\lambda_1$ is the linear loss rate constant and  $\lambda_2$ is the non-linear loss rate constant.  Substituting Eq.~(\ref{eq:lambda}) into Eq.~(\ref{eq:dtLPT1}) gives a rate equation with the same form as the temperature-limited MOT loading rate Eq.~(\ref{eq:dtMOT}),
\begin{equation}
\frac{dN_I}{dt} = L_I - \lambda_1 N_I - \lambda_2 N_I^2.
\label{eq:dtLPT2}
\end{equation}

We found that the solution to Eq.~(\ref{eq:dtLPT2}) fit the time dependent loading data slightly better than the fits shown in Fig.~\ref{fig:LoadN1}, probably because of the additional fitting parameter.   Because the fits were slightly better, we used the steady-state ion population fitting results from the solution to rate Eq.~(\ref{eq:dtLPT2}) as the independent variable in Fig.~\ref{fig:giaNI}.  However, we found that there was little to no difference in the fit results of $L_I$ or $\tilde{N}_I$ when we used the solution to rate Eq.~(\ref{eq:dtLPT1}) vs.~that of Eq.~(\ref{eq:dtLPT2}). The uncertainty in the steady-state values comes from propagating the uncertainty in the ion loading fit results and the CEM calibration fit results.

Unfortunately, we find that the LPT loss rates  $\lambda_1$ and  $\lambda_2$ still have an $I_\mathrm{pi}$  dependence, which suggests that Eq~(\ref{eq:dtLPT2}) is also not the correct rate equation model for LPT saturation.
\vspace{-1em}
\subsubsection{LPT decay}
\label{sec:LPTdecay}
\begin{figure}[t]
   \centering
   \includegraphics[width=3.25 in]{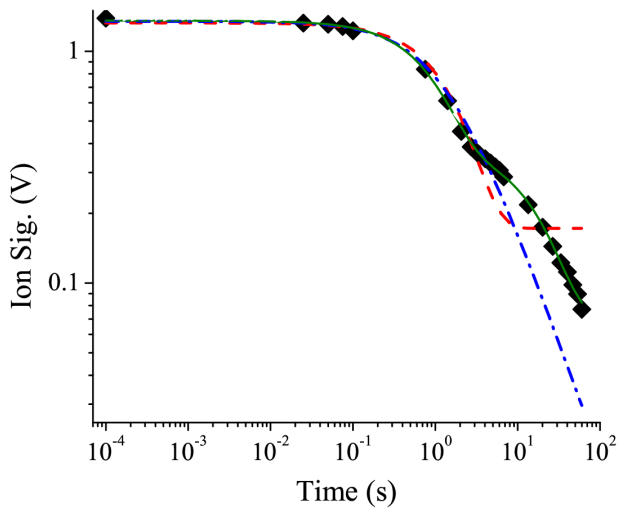}
   \caption{(Color online). Experimental LPT ion population decay as a function of time on a log-log scale.  The LPT is saturated with the type I MOT and  $I_\mathrm{pi} \approx 670$ mW/cm$^2$. Loading from the type II MOT gives qualitatively identical behavior. The decay is initially nonexistent and is then followed by a sudden rapid decay (after 0.1 s) and then a slower decay (after 1.1 s). We have fit the decay curve to three solutions: the red dashed curve is the solution to Eq.~(\ref{eq:dtLPT1}) with $L_I = 0$, the blue dot-dashed curve is the solution to Eq.~(\ref{eq:dtLPT2}) with $L_I =0$, and the solid green curve is for a double exponential decay. The uncertainty in the data is smaller than the plot markers.}
   \label{fig:ExpDec}
\vspace{-0.75em}
\end{figure}

\vspace{-1em}
The missing piece to the LPT loading dynamics is how to accurately model the LPT loss rate. In an attempt to better understand the loss mechanism, we looked at the LPT ion decay when the LPT is initially saturated, as seen in Fig.~\ref{fig:ExpDec}.  Experimentally, the LPT is initially saturated with either the type I or II MOT.  After loading to saturation, the MOT is turned off and the ions are held in the trap for some delay without the MOT.  The ions are then extracted and detected with the CEM. The process is repeated for increasing delay times until only a small ion signal is detected.

We did not see the simple exponential decay (red dashed curve in Fig.~\ref{fig:ExpDec}) often observed \cite{Green:2007, Grier:2009, Sullivan:2011}, even previously by our own group \cite{Sivarajah:2012}.  We also found poor agreement with the decay model developed in Ref.~\cite{Ravi:2012}. We found a slight improvement when the ion decay was fit to the solution of Eq.~(\ref{eq:dtLPT2}) with $L_I=0$ (blue dot-dash curve in Fig.~\ref{fig:ExpDec}), but the best fit was with a two exponential decay (green solid curve in Fig.~\ref{fig:ExpDec}).  Granted, the two exponential decay has the largest number of free parameters, but this equation emphasizes that there is an initial rapid loss after 0.1 s and then a more gradual loss after 1.1 s. We suspect that the departure from the simple exponential decay is due to the fact that the LPT is saturated, which was not the case in Ref.~\cite{Sivarajah:2012}.

Saturation of the LPT occurs when the Coulomb space charge force and the spring force are balanced. Therefore it is reasonable to expect space charge effects to play some role at saturation, even in the weakly coupled regime, $\Gamma \ll 1$. The initial rapid loss may be due to space charge effects playing a significant role in the dynamics or possibly caused by a slightly non-gaussian LPT saturated spatial distribution. The two exponential decay solution would come from a second order rate equation, like that of an un-driven overdamped harmonic oscillator. Unfortunately, we do not have a physical motivation for introducing such a rate equation at this time.

We have also conducted simulations of 500 interacting ions in an idealized quadrupole field, which decay from an ion trap over 100,000 rf periods. The ions are simulated in the absence of any ion-neutral collisions, patch fields, or electrode imperfections. An ion is considered lost from the trap when its position exceeds a critical radius. In a real LPT the critical radius would be determined by the effective ion trap depth or physical edge of the trapping electrodes (whichever is smallest).  However, in the simulations the critical radius is arbitrarily chosen to approximately be double the size of the initial simulated ion cloud width.
\begin{figure}[t]
   \centering
   \includegraphics[width=3.25 in]{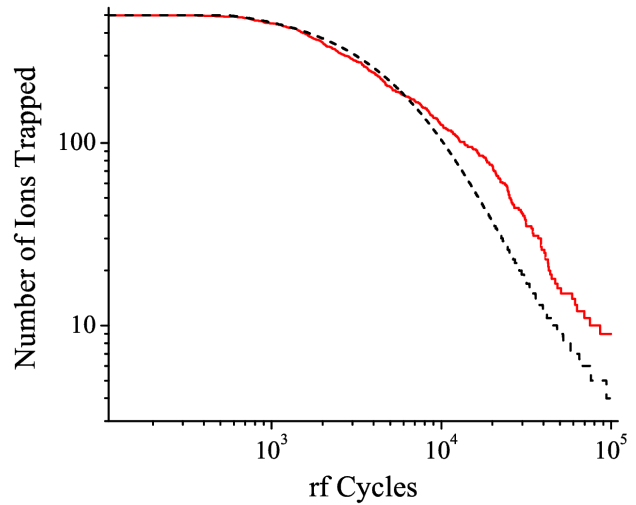}
   \caption{(Color online). Ion trap simulations (red solid curve) and numerical diffusing Gaussian model (black dashed curve) of 500 ions decaying from an idealized ion trap.  The simulation and model calculations show good quantitiative agreement with each other and good qualitative agreement with the experimental data in Fig.~\ref{fig:ExpDec}.}
   \label{fig:SimDec}
\vspace{-0.75em}
\end{figure}

The ion trap simulations (shown as the solid red curve in Fig.~\ref{fig:SimDec}) also exhibit three regimes: a brief initial period of stability, then a rapid loss followed by a more gradual loss at low ion number.  The ion trap simulations show good qualitative agreement despite being idealized.  Furthermore, the simulated trap does not represent the exact dimensions of our LPT, as they are not the same \textsc{simion} simulations discussed earlier. The qualitative agreement suggests that ion-ion rf heating \cite{Nam:2014,Blumel:1988,Ryjkov:2005,Zhang:2007} is the main cause of the trap decay, since it is the only simulated loss mechanism here. In the interest of having reasonable computation times, the simulations are performed with much higher ion cloud densities and within a much smaller trap depth, as compared to the actual experimental conditions.  Therefore,  Fig.~\ref{fig:SimDec}'s decay occurs over a much shorter period of time, making the comparison strictly qualitative.

We also performed numerical simulations of a likely physical underlying process -- an ion cloud with a Gaussian spatial distribution diffuses due to ion-ion rf~heating, until one of the hot ions in the tail reaches the critical radius and is lost from the trap. When an ion is lost, it carries off a fraction of the cloud's energy lowering the temperature of the cloud. This causes a shrinking of the cloud, which diffuses back to the critical radius. As the number of ions is reduced, each successive ion removes a larger fraction of the total energy when it is lost, leading to a non-exponential decay, as seen in the black dashed curve in Fig.~\ref{fig:SimDec}.  The diffusing Gaussian model shows good quantitative agreement with the ion trap simulations. 

In this section we have revised the loading model from Ref.~\cite{Lee:2013}.  In doing so, we are confident that we can accurately model the LPT loading rate $L_I$ and the steady-state ion population $\tilde{N}_I$, but have yet to determine a completely satisfactory closed-form analytic solution to both the LPT loading and decay rate equations. We plan to continue our studies on the subject of LPT loading and decay and we plan to present our findings in a more detailed manuscript in the near future. 
%%%%%%%%%%%%%%%%%%%%%%%%%%%%%%%%%%%%
\vspace{-1.35em}
\subsection{Dark \ce{Na+} ion cloud size}
\label{sec:CloudSize}
\vspace{-1em}
To determine $k_\mathrm{ia}$ we must first determine the dark \ce{Na+} ion cloud size.  For optically accessible ion clouds this can be accomplished by simply imaging the ion cloud in the same way we image the MOT, but for a dark ion cloud this is not an option.  In principle, if the trap is saturated and the radial trap depth $D$ is known then the maximum transverse radius of the ion cloud $\tilde{r}_I$ can be determined by equating the depth to the harmonic potential energy of the outermost ion's turning point, which gives
\begin{equation}
\tilde{r}_{I,1} = \sqrt{\frac{2D}{\omega_r^2}}
\label{eq:rmax}
\end{equation} 
in the radial dimension \cite{Smith:2014,Lee:2013, Ray:2014}.  Because the radial depth is much greater than the axial trap depth \cite{Raizens:1992}, we can assume the cloud is limited by the equally partitioned \cite{Berkeland:1998} transverse secular energy mode, making the maximum axial extent $\tilde{r}_{I,3} = \omega_r \tilde{r}_{I,1}/\omega_a$, if we assume a harmonic axial potential. However, it is difficult to experimentally determine the effective trap depth, which can be quite different from the theoretical single ion idealized quadrupole radial trap depth \cite{Raizens:1992}, which is merely a function of the trap voltage settings.  For example, this was found to be the case in Ref.~\cite{Ravi:2012}.
\begin{figure}[t]
   \centering
   \includegraphics[width=3.25 in]{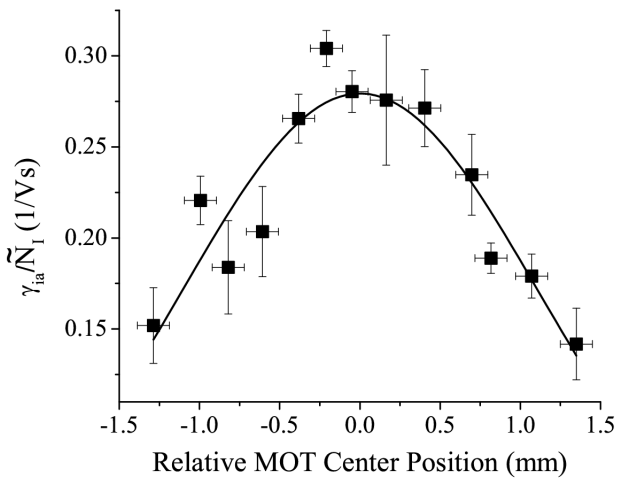}
   \caption{(Color online). MOT ion-atom loss rate $\gamma_\mathrm{ia}$ normalized by the steady-state ion population $\tilde{N}_I$ as a function of the center position of the type I MOT (as measured with the CMOS camera) relative to the geometric center of the LPT [$x_{0,1}$ in Eq.~(\ref{eq:C})].  The data are fit to Eq~(\ref{eq:gia}) and the fitted ion cloud radius is $r_{I,1} = 1.6 \pm 0.1$ mm.}
   \label{fig:MOTPos}
\vspace{-0.75em}
\end{figure}

The first upper bound on the radial extent of the ion cloud is the mechanical inner electrode radius of the trap $r_0 \approx 9.5$ mm, as seen in Fig.~\ref{fig:MOTs}.  We can reduce this upper bound by using our \textsc{simion} simulations. We simulated an ion that is initialized with no initial kinetic energy at ever increasing transverse displacement from the LPT's nodal line \cite{Ray:2014}.  If the ion starts at a distance $\ge 3$ mm from the nodal line at the experimental trap settings, we find that the ion cannot remain trapped for more than two secular periods. If we consider this upper bound to be equivalent to the $1/e^2$ radius of the Gaussian distribution, then the upper bound on $r_{I,1}=3/\sqrt{2} \approx 2.12$ mm.

Other groups \cite{Schmid:2010,ZipkesPRL:2010,ZipkesNature:2010} have used a single trapped ion in a hybrid trap to probe a neutral BEC. We have essentially employed the reverse process -- we use the MOT to probe a dark ion cloud. By translating the MOT across the saturated ion cloud along one transverse dimension, we measured $\gamma_\mathrm{ia}$ as a function of the changing concentricity function $C \left ( x_{0,1} \right )$ in Eq.~(\ref{eq:C}).  As we translate the MOT the steady-state number of ions changes slightly, since the PI rate changes slightly as well as the temperature of the saturated ion cloud.  Therefore, we normalize $\gamma_\mathrm{ia}$ to the steady-state ion population point for point.  We found the normalized ion-atom loss rate fit well to Eq.~(\ref{eq:gia}), as seen in Fig.~\ref{fig:MOTPos}, which supports our claim that the ion cloud had a Gaussian spatial distribution.

We have assumed that as the MOT is translated $k_\mathrm{ia}$ remains constant. Because the temperature of the ion cloud will change when the LPT is loaded from a MOT displaced off the nodal line, $k_\mathrm{ia}$ is technically different from point to point.  However, since $k_\mathrm{ia}$ has a weak temperature dependence the model still fits well.

Measurements taken over several days found that the saturated ion cloud size did not depend on the PI intensity used.  Typical fit results gave $r_{I,1} = 1.6 \pm 0.1$ to $r_{I,1} = 1.9 \pm 0.1$ mm, always less than but close to the simulation upper bound.  Therefore, we will use the experimental data as a lower bound on the ion cloud radius of $r_{I,1} = 1.75$ mm.

Instead of using the ratio of the secular periods to determine the lower and upper bound on the axial extent of the ion cloud, we used our \textsc{simion} simulations with an ion initialized at the center of the trap having the kinetic energy equivalent to the potential energy at the maximum radial turning point for $r_{I,1} = 1.75$ mm and $r_{I,1} = 2.12$ mm, respectively. The \textsc{simion} simulations found the lower and upper bound axial extent to be $r_{I,3} = 10.57$ mm and $r_{I,3} = 12.10$ mm, respectively.  Using the \textsc{simion} simulations should be more accurate because it models the actual LPT electrode geometry, which yields a more quartic axial electrical potential than harmonic.

Having determined the ion cloud size and the MOT dimensions via CMOS camera measurements, $\tilde{V}_\mathrm{ia}$ can be determined.  The final results are summarized in Table~\ref{tab:results}. The experimental $k_\mathrm{ia}$ is calculated using the slopes from Fig.~\ref{fig:giaNI} while the theoretically determined $k_\mathrm{ia}$ values are weighted averages of the values found in Table~\ref{tab:kia}, based on the MOT $f_e$.
\begin{table}[t]
\vspace{-0.5em}
\caption{Table of total rate coefficient experimental and theoretical results for the type I and type II MOTs.  The saturated ion-atom volume $\tilde{V}_\mathrm{ia}$ is determined using Eq.~(\ref{eq:Via}) with input from measurements discussed in Sec.~\ref{sec:CloudSize}.}
\centering
\begin{ruledtabular}
\begin{tabular}{c|c|c|c}
\multirow{2}{*}{MOT} & \multirow{2}{*}{$\tilde{V}_\mathrm{ia}$ (cm$^3$)} & Experimental & Theoretical \\
& & $k_\mathrm{ia} \left ( \mathrm{cm}^3/\mathrm{s} \right )$ & $k_\mathrm{ia} \left ( \mathrm{cm}^3/\mathrm{s} \right )$ \\
\hline
type I & $ 0.247 \pm 0.061 $ & $7.4 \pm 1.9 \times 10^{-8}$ & $8.62 \pm 0.07 \times 10^{-8}$ \\
type II & $ 0.267 \pm 0.060 $ & $1.10 \pm 0.25\times 10^{-7}$ & $1.14  \pm 0.01 \times 10^{-7}$ \\
\end{tabular}
\end{ruledtabular}
\label{tab:results}
\vspace{-1.25em}
\end{table}
%%%%%%%%%%%%%%%%%%%%%%%%%%%%%%%%%%%%
%%%%%%%%%%%%%%%%%%%%%%%%%%%%%%%%%%%%
\vspace{-1.5em}
\section{CONCLUSION}
\label{sec:Con}
\vspace{-1em}
We have demonstrated a modified version of a method, originally reported in Ref.~\cite{Lee:2013} for \ce{Rb+}-- \ce{Rb}, for measuring the total ion-atom collision rate coefficient of \ce{Na} on optically dark \ce{Na+}.  The experimental results show very good agreement with previously reported fully quantal \textit{ab intio} calculations. In determining $k_\mathrm{ia}$ we demonstrated that the MOT can be used as a probe of a dark ion cloud spatial distribution. We have also measured the two-body \ce{Na} MOT atom-atom collision rate coefficient $\beta$ and the PI cross section $\sigma_\mathrm{pi}$ at 405 nm radiation for both the type I and II MOTs.  The measurements of $\beta$ and $\sigma_\mathrm{pi}$ showed good agreement with previously reported experimental and theoretical values.

For optically bright ion clouds, the charge-exchange rate coefficient can be determined by the ion decay alone.  However, by also using the MOT ion-atom loss rate to determine the total collision rate coefficient, the elastic scattering rate coefficient can be determined by subtracting the two results.  We plan to implement this procedure in measurements on the \ce{Ca+}-- \ce{Na} system. Finally, we have presented some preliminary simulation and experimental results toward the development of an analytical closed-form model of LPT trap loss, saturation, and loading dynamics. 
%%%%%%%%%%%%%%%%%%%%%%%%%%%%%%%%%%%%
%%%%%%%%%%%%%%%%%%%%%%%%%%%%%%%%%%%% 
\vspace{-1.5em}
\section{Acknowledgments}
\label{sec:Acknowledgments}
\vspace{-1em}
We would like to acknowledge support from the NSF under Grant No.~PHY-1307874. We thank Jian Lin, Oleg Makarov, Kristen Basiaga, Charles Talbot, and Ilamaran Sivarajah for their preliminary work on the hybrid trap project.  We would also like to thank our University of Connecticut theoretical collaborators Robin C$\hat{\mathrm{o}}$t$\acute{\mathrm{e}}$, Harvey Michels, and John Montgomery.
%%%%%%%%%%%%%%%%%%%%%%%%%%%%%%%%%%%%
%%%%%%%%%%%%%%%%%%%%%%%%%%%%%%%%%%%%
%merlin.mbs apsrev4-1.bst 2010-07-25 4.21a (PWD, AO, DPC) hacked
%Control: key (0)
%Control: author (8) initials jnrlst
%Control: editor formatted (1) identically to author
%Control: production of article title (-1) disabled
%Control: page (0) single
%Control: year (1) truncated
%Control: production of eprint (0) enabled
%
\end{document}